\documentclass[twocolumn,pre,superscriptaddress,floatfix,longbibliography]{revtex4-1}

\usepackage{graphics}
\usepackage{graphicx}
\usepackage{amsmath}
\usepackage{amssymb,xcolor}
\usepackage{gensymb}
\usepackage{ulem}

\definecolor{grisclair}{rgb}{0.6,0.6,0.6}

\newcommand{\beq}{\begin{equation}}
\newcommand{\ee}{\end{equation}}

\begin{document}

\title{Superstability of micrometer viscous jets surrounded by weakly-elastic viscous shells}

\author{A. Rubio-González}
\address{Depto.\ de Ingenier\'{\i}a Mec\'anica, Energ\'etica y de los Materiales and\\ 
Instituto de Computaci\'on Cient\'{\i}fica Avanzada (ICCAEx),\\
Universidad de Extremadura, E-06006 Badajoz, Spain}
\author{D. Fernández-Martínez}
\address{Depto.\ de Ingenier\'{\i}a Mec\'anica, Energ\'etica y de los Materiales and\\ 
Instituto de Computaci\'on Cient\'{\i}fica Avanzada (ICCAEx),\\
Universidad de Extremadura, E-06006 Badajoz, Spain}
\author{M. Vakili}

\address{European XFEL, Holzkoppel 4, 22869 Schenefeld, Germany}
\author{J. M. Montanero}
\address{Depto.\ de Ingenier\'{\i}a Mec\'anica, Energ\'etica y de los Materiales and\\ 
Instituto de Computaci\'on Cient\'{\i}fica Avanzada (ICCAEx),\\
Universidad de Extremadura, E-06006 Badajoz, Spain}

\begin{abstract}
Our experiments demonstrate the strong stabilizing effect of a viscoelastic shell on flow-focused viscous jets. The results suggest that stabilization occurs when the polymer viscosity, $\eta_p$, exceeds a critical threshold, provided the extensional relaxation time, $\lambda$, is large enough to allow the polymer molecules to undergo the coil–stretch transition. Under these conditions, substantial extensional viscoelastic stresses develop as the liquid accelerates. Since both $\eta_p$ and $\lambda$ increase with shell viscosity, the polymer concentration required to stabilize the viscous jet decreases accordingly. Remarkably, moderately to highly viscous jets can be stabilized with very small polymer concentrations, for which no beads-on-a-string (BOAS) instability is observed.
\end{abstract}

\maketitle

\section{Introduction}

Gaseous flow focusing \citep{G98a,DWSWSSD08} is a gas-liquid tip-streaming flow that typically operates in microjetting mode, enabling the ejection of jets with diameters on the order of a few microns and speeds exceeding 100 km/h by purely hydrodynamic means. This technique has been widely used in different technological fields, such as pharmacy and materials engineering \citep{M24}.

The X-ray free-electron laser (XFEL) produces highly energetic, coherent X-ray pulses with femtosecond duration. These machines have opened new avenues for studying the structure and dynamics of condensed matter, liquids, and gases, spanning fields from planetary science and atomic physics to structural biology. \citet{Cetal11} first introduced hydrated protein crystals to the X-ray beam, in a vacuum, using flow focusing. Since then, flow-focused liquid jets have been used in many serial femtosecond crystallography (SFX) experiments at XFEL facilities worldwide \citep{W18}. 

Shear and extensional rates of the order of $10^6$ s$^{-1}$ are reached in the inception region of a flow-focused jet. When a small amount of polymer is dissolved in the liquid, the coil-stretch polymer transition occurs even for polymer relaxation times of the order of tens of microseconds \citep{RVGM22}. This sharply increases the liquid extensional viscosity. Stretched polymers somehow entangle within the micrometer-scale jets \citep{V26}, delaying extensional-viscosity relaxation on a time scale much longer than the polymer relaxation time. This results in superstable jets flying distances thousands of times their diameters \citep{RGVMC22}.

\citet{RMVKBCG25} recently used a viscoelastic shell to achieve greater control and stability of the flowing jet. They maintained a buffer formulation within the jet core. The core was enveloped by a shell of miscible, low-concentration aqueous solution containing a biocompatible polymer. This considerably extended the jet length while reducing the liquid flow rate by an order of magnitude \cite{RGVMC22,RVGM22}. \citet{V26} demonstrated the effectiveness of this method for delivering protein microcrystals in buffer compositions of various viscosities for SFX.

Under certain experimental conditions, the jets stabilized by the aqueous shell suffer from a type of beads-on-a-string (BOAS) instability \citep{RVGM22,V26}. This can be viewed as the viscoelastic equivalent of the Rayleigh instability \citep{EV08}, in which the jet does not break up until very late times but forms beads connected by thin filaments due to the elasticity \citep{EY84}. The BOAS instability may pose a serious obstacle in X-ray diffractive imaging experiments because the small beads emerging along the filament lead to the so-called blistering instability \cite{OM05} and may reduce sample-beam overlap. Furthermore, the beads represent additional background signal, which may hinder data collection from weekly-scattering (non-crystalline) samples \citep{BRM23,WE24,Vakili2026_2}. The presence of beads also restrict the unperturbed section of the liquid column to the vicinity of the nozzle tip and therefore reduce achievable laser pump–probe time delays in time-resolved studies \citep{V26}. 

This paper explores the effect of shell viscosity on jet stabilization across different jet viscosities. We will show that viscous shells require lower polymer concentrations to achieve the same stabilizing effect and do not suffer from the BOAS instability. Our results indicate that stabilization occurs when the polymer viscosity exceeds a critical threshold, provided that polymer relaxation time is sufficiently large to trigger the polymer coil–stretch transition.

\section{Experimental Method}
\label{met}

\begin{figure}[!]
\begin{center}
\resizebox{\columnwidth}{!}{\includegraphics{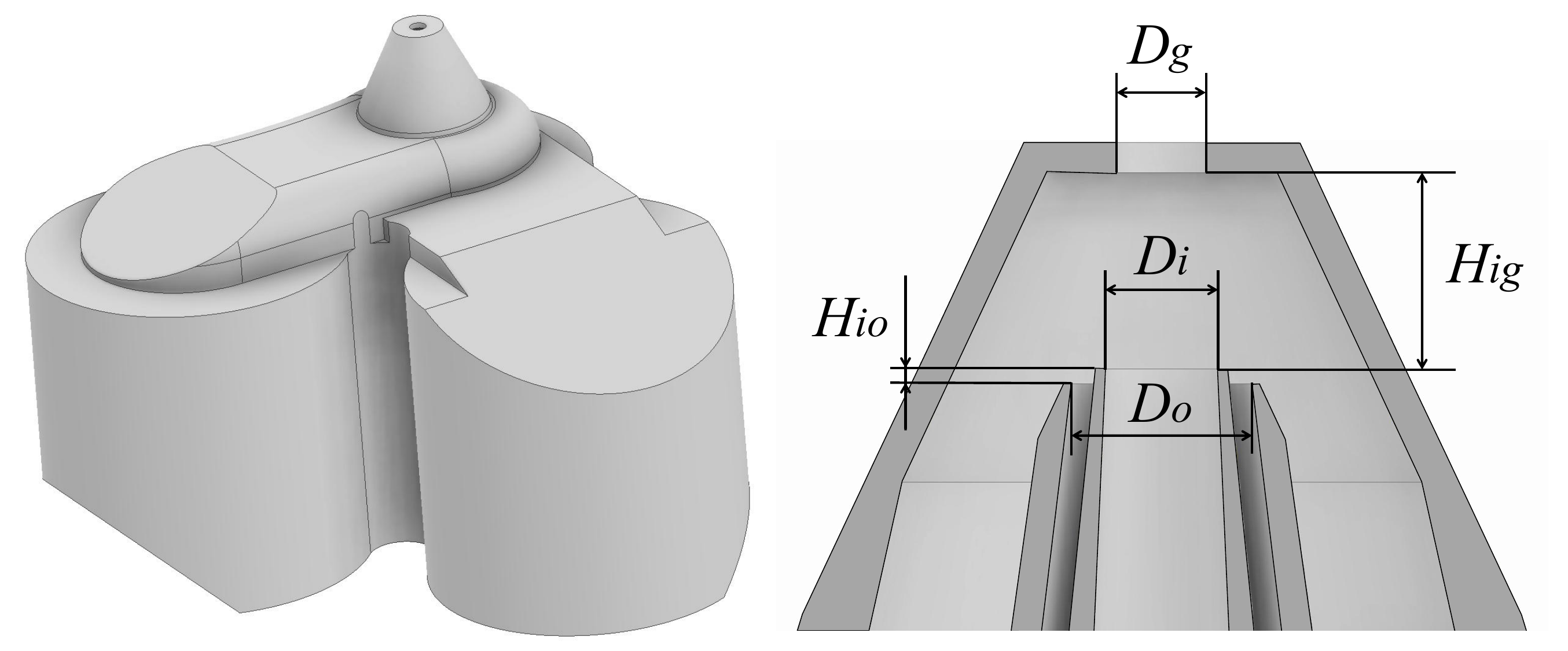}}
\end{center}
\caption{Screenshot of the nozzle with indications of the design parameters.}
\label{geom}
\end{figure}

We printed the gaseous flow-focusing ejector using the Nanoscribe Photonic Professional GT2 with the Dip-in Laser Lithography (DiLL) configuration, a 25$\times$ objective lens, and the IP-S photoresist. It consisted of a sharp two-fold concentric capillary coaxially placed inside a converging nozzle (Fig.\ \ref{geom}). We conducted experiments using nozzle labeled as N9 in Ref.\ \citep{RGVMC22}, with the following characteristic lengths: $D_i=75$ $\mu$m, $D_o=120$ $\mu$m, $D_g=60$ $\mu$m, $H_{ig}=130$ $\mu$m, and $H_{io}=10$ $\mu$m. This nozzle has been shown to produce superstable protein microcrystal jets surrounded by the aqueous viscoelastic shell \citep{RGVMC22}. 

The focused liquid was injected across the inner capillary at a constant flow rate $Q_i$ with a syringe pump ({\sc Legato}210, KD-{\sc Scientific}). The fluid shell was injected through the outer capillary at a constant flow rate $Q_o$ with another syringe pump ({\sc Legato}210, KD-{\sc Scientific}). We set the upstream stagnation pressure of the focusing air current to $p_0=1.5$ bar using a pressure regulator on the compressed-air line. We mounted the ejector on the cap of a discharge glass cell and set the pressure inside the cell to $p_e=42$ mbar (the lowest pressure achievable with our experimental setup) using a suction pump. Under these conditions, we estimated the air mass flow rate to be $m'_g=81$ mg/min.

We used a high-speed video camera ({\sc Fastcam Mini} UX50) to take digital microscopy images of the jet at $4000$ fps. The camera was equipped with optical lenses ({\sc Optem Zoom 70XL}) and a microscope objective ({\sc Optem} 10X). The images consisted of 974$\times$768 pixels. The magnification was 3×, resulting in 0.444 $\mu$m/pixel. We illuminated the jet with a LED (KL 2500, {\sc Schott}). To examine jet behavior at different jet stations, the camera was displaced vertically using a triaxial translation stage with its vertical axis motorized ({\sc Thorlabs} Z825B) and controlled by the computer. We analyzed the images to measure jet diameter in pixels at different jet stations. For some representative cases, we repeated the experiment several times. We concluded that the experimental uncertainty in the jet's diameter was approximately one pixel. 

Experiments were conducted using aqueous polyethylene glycol (PEG) solutions with the molecular weight $M_w=6\times 10^3$ g $\text{mol}^{-1}$ (PEG6K) at the concentration 6.5\% (wt) (Buffer 1), 15\% (wt) (Buffer 2), 32\% (wt) (Buffer 3), and 40\% (wt) (Buffer 4). Buffer 2 is very similar to the liquids used to carry microcrystals in SFX. Buffer 1 has an intermediate viscosity between that of Buffer 2 and the water used in previous experiments \citep{RRCHGM21,RGVMC22}. Finally, we increased the PEG concentration in Buffer 3 and Buffer 4 to explore the behavior of a highly viscous jet. We added polyethylene oxide (PEO) with the molecular weight $M_w=100\times 10^3$ g$\text{mol}^{-1}$ (PEO100K) to the liquid shell to stabilize the jet.

Table \ref{liquids} shows the solvent properties and the viscoelastic solution properties at the optimal PEO100K concentration (see Sec.\ \ref{res}). The table also shows the properties of the aqueous PEO100K solution at the optimal concentration determined by \citet{RRCHGM21}. The density was measured with a Gay-Lussac pycnometer ({\sc Glassco}) of $10\pm1$ ml and a precision balance. We measured the solution viscosity $\eta$ and the PEO100K contribution $\eta_p$ to that viscosity with a Cannon-Fenske viscometer ({\sc Fungilab}). We did not observe shear thinning. Therefore, the buffer fluids and the viscoelastic solutions can be regarded as Newtonian and Boger liquids \citep{J09}, respectively. We determined the extensional relaxation time $\lambda$ using the slow retraction method \citep{SVSMA17}. Adding PEO100K produces a weak-to-moderate viscoelastic effect ($\lambda\sim 10-10^3$ $\mu$s).  The extensional relaxation time $\lambda$ increases significantly with solvent viscosity, although the PEO100K concentration decreases. Finally, we measured the surface tension using the TIFA-AI method \citep{FMC07}. 

\begin{table*}
\begin{tabular}{ccccccc}
Liquid & $\rho$ (kg/m$^3$) & $\eta$ (mPa·s) & $\eta_p$ (mPa·s) & $\lambda$ ($\mu$s)& $\sigma$ (mN/m) & $\langle dw/dz\rangle^{-1}$ ($\mu$s)\\
\hline
Water                   & 998.5  & 1.00 &-  &-  & 71.8 & -\\
Water+PEO100K 0.75\%       & 1000.0   & 1.89   &$0.89$  &$20$ & 60.0 & 17 \\
Buffer 1                & 1016.6 & 2.49 & - & - & 61.7 & -\\
Buffer 1+PEO100K 0.4\%  & 1016.5 & 3.21 & $0.72$ & $45$ & 61.0 & 23\\
Buffer 2                & 1061.8 & 6.71 & - & - & 59.6& - \\
Buffer 2+PEO100K 0.18\% & 1062.3 & 7.57& $0.86$ & $126$ & 59.7& 22\\
Buffer 3                & 1057.2 & 35.62& - & - & 58.5& -\\
Buffer 3+PEO100K 0.045\% & 1052.5 & 36.49 & $0.87$ & $1276$ & 57.9 & 21\\
Buffer 4                & 1067.7 & 65.50 & - & - & 58.7 & -\\
Buffer 4+PEO100K 0.03\% & 1067.1 & 66.43& $0.93$ & $2045$& 58.9 & 17\\
\hline
\end{tabular}
\caption{Properties of liquids used in the experiments. The table also shows the inverse of the estimated average strain rate in the liquid meniscus, $\langle dw/dz\rangle=v_m/l_m$, where $v_m$ and $l_m$ are estimates of the meniscus velocity and length, respectively.}
\label{liquids}
\end{table*}

\section{Results and Discussion}
\label{res}

\begin{figure}[!]
\begin{center}
\resizebox{0.75\columnwidth}{!}{\includegraphics{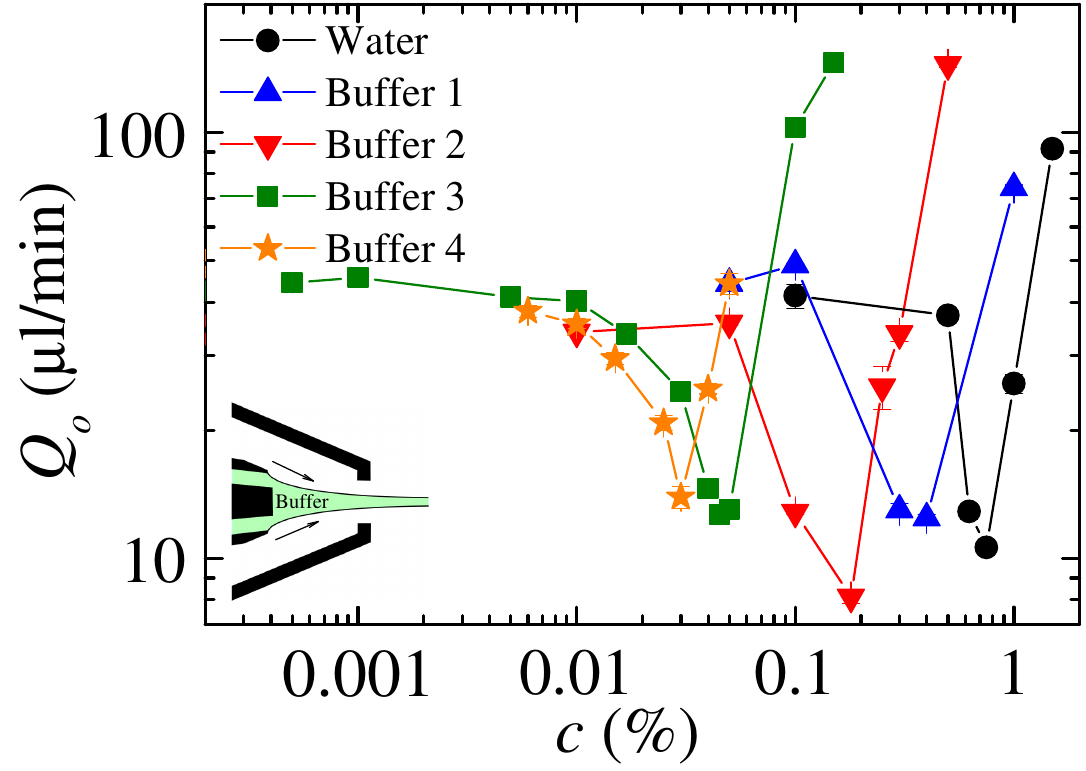}}
\end{center}
\caption{Outer flow rate $Q_o$ at the stability limit as a function of the the PEO100K concentration for $Q_i=0$ and a mass flow rate of $m'_g=81$ mg/min.}
\label{con}
\end{figure}

\begin{figure}[!]
\begin{center}
\resizebox{0.75\columnwidth}{!}{\includegraphics{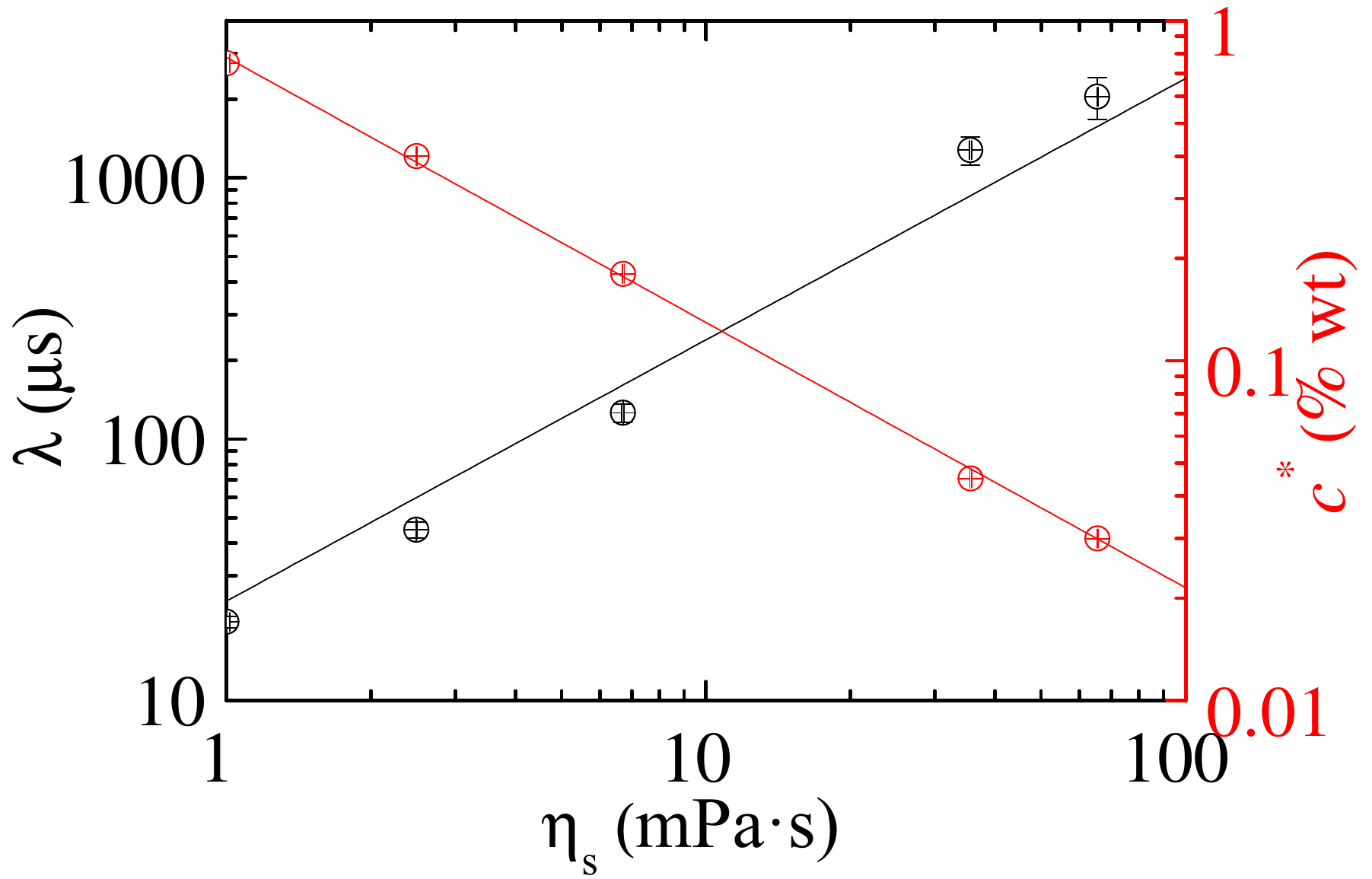}}
\end{center}
\caption{Polymer relaxation time $\lambda$ (left) and optimal concentration $c^*$ (right) as a function of the solvent viscosity $\eta_{s}$. The lines are the fits $\lambda=24 \eta_s$ and $c^*=0.78\eta_s^{-0.78}$ to the experimental data.}
\label{lambda}
\end{figure}

First, we analyzed the effect of the polymer concentration for all the liquids considered in this work. Specifically, we determined the minimum stable outer flow rate $Q_o$ for $Q_i=0$ (Fig.\ \ref{con}). As \citet{RGVMC22} explain, the polymer coil-stretch transition is triggered above a threshold polymer concentration because of the extremely high shear and extensional rates reached at the tapering meniscus tip. This explains the sharp decrease in the minimum value of $Q_o$. Pull-out instability  (the tapering meniscus detaches from the feeding capillary edge) \citep{WHBMB12} destabilizes jet emission at higher concentrations, which explains the increase in the minimum value of $Q_o$. We determined the optimal PEO100K concentration, $c^*$, to be the one that yields the minimum stable outer flow rate $Q_o$. We selected concentrations of 0.4\% (wt), 0.18\% (wt), 0.045\% (wt), and 0.03\% (wt) for Buffer 1, Buffer 2, Buffer 3, and Buffer 4, respectively (Fig.\ \ref{lambda}). In our experiments, solvent viscosity enhances the elastic effect of PEO100K ($\eta_p$ and $\lambda$ increase with solvent viscosity for a given polymer concentration). Therefore, the optimum PEO100K concentration decreases as solvent viscosity increases. However, the minimum value of $Q_o$ is essentially the same in all the cases (Fig.\ \ref{con}).

It is natural to hypothesize that polymer stress stabilizes the flow for sufficiently large values of $\tau^p_{zz}/\tau^s_{zz}$, where $\tau^s_{zz}\simeq \eta_s dw/dz$ is the solvent viscous stress. Consider the upper-convected Maxwell model \citep{BAH87} combined with the 1D approximation \citep{E97}, leading to the following equation for the axial polymeric stress $\tau^p_{zz}$ \citep{YPR14}:
\begin{equation}
w\frac{d\tau^p_{zz}}{dz}=2\frac{dw}{dz}\tau^p_{zz}+\frac{2\eta_p}{\lambda}\frac{dw}{dz}-\frac{\tau^p_{zz}}{\lambda},    
\label{relax}
\end{equation}
where $w(z)$ is the axial velocity. If one assumes that $dw/dz\simeq\langle dw/dz\rangle\gg \lambda^{-1}$ in the meniscus, the integration of (\ref{relax}) within that region approximately gives \citep{HYR08,YPR14}
\begin{equation}
\label{rel}
\tau^p_{zz}=\tau^{p0}_{zz}\exp\left[2\,\langle dw/dz\rangle\, (t-t_0)\right],
\end{equation}
where $\tau^{p0}_{zz}$ is the axial polymer stress at the entrance of the liquid meniscus ($t=t_0$). The stress $\tau^{p0}_{zz}={\cal F}\lambda\eta_p Q_o^2$ builds up at the walls of the nozzle outer channel due to the steady Poiseuille flow, where ${\cal F}={\cal F}(D_i,D_o)$ is a constant independent of the liquid properties. Our results show that $\lambda\sim \eta_s$ (see Fig.\ \ref{lambda}) and $\langle dw/dz\rangle\simeq \text{const}$ (see Table \ref{liquids}), and therefore $\tau^p_{zz}/\tau^s_{zz}$ is proportional to $\eta_p$ and independent of $\lambda$. 

Interestingly, the optimum concentration was obtained over a narrow range of $\eta_p$ ($0.7\lesssim \eta_p\lesssim 0.9$ mPa$\cdot$s) and across a wide range of $\lambda$ ($20\lesssim \lambda\lesssim 2000$ $\mu$s), even though the average strain rate is commensurate for all the liquids  (Table \ref{liquids}). As our analysis predicts, stabilization occurs when $\eta_p$ exceeds a threshold, provided $\lambda$ exceeds the value required for the polymer-coil stretch transition. In this case, liquid acceleration builds up sufficiently large extensional stresses to stabilize the flow. Fewer PEO100K molecules produce insufficient extensional thickening, while more molecules generate viscoelastic stresses that cause pull-out instability.

The behavior described above differs substantially from that observed in experiments with aqueous PEO solutions of different molecular weights flow-focused through a simple nozzle \citep{RGVMC22}. In that case, adding a small number of long molecules to the low-viscosity solvent introduces a characteristic time (the extensional relaxation time, $\lambda$) that matches the hydrodynamic time of the problem (the inverse of the strain rate), producing a ``nonlinear resonant behavior" leading to flow stabilization \citep{RGVMC22}. The polymer viscosity seems to play a secondary role in those experiments. Polymers are carried by a low-viscosity solvent across the (simple) nozzle, so they are essentially relaxed at the entrance of the tapering meniscus, which may explain the difference with respect to the behavior observed in our experiments.

PEG6K and PEO100K have the same chemical composition. The higher molecular weight of PEO100K stabilizes the flow even though its mass concentration is much smaller than that of PEG6K. For instance, the Buffer 4 flow is stabilized even though the PEO100K mass
concentration is approximately 1300 times lower than that of PEG6K. This implies approximately 1 molecule of PEO100K per 22,000 molecules of PEG6K. The coil-stretch transition of the PEO100K molecules arises from hydrodynamic interactions among them, despite their low concentration.

\begin{figure*}[!]
\begin{center}
\resizebox{1.2\columnwidth}{!}{\includegraphics{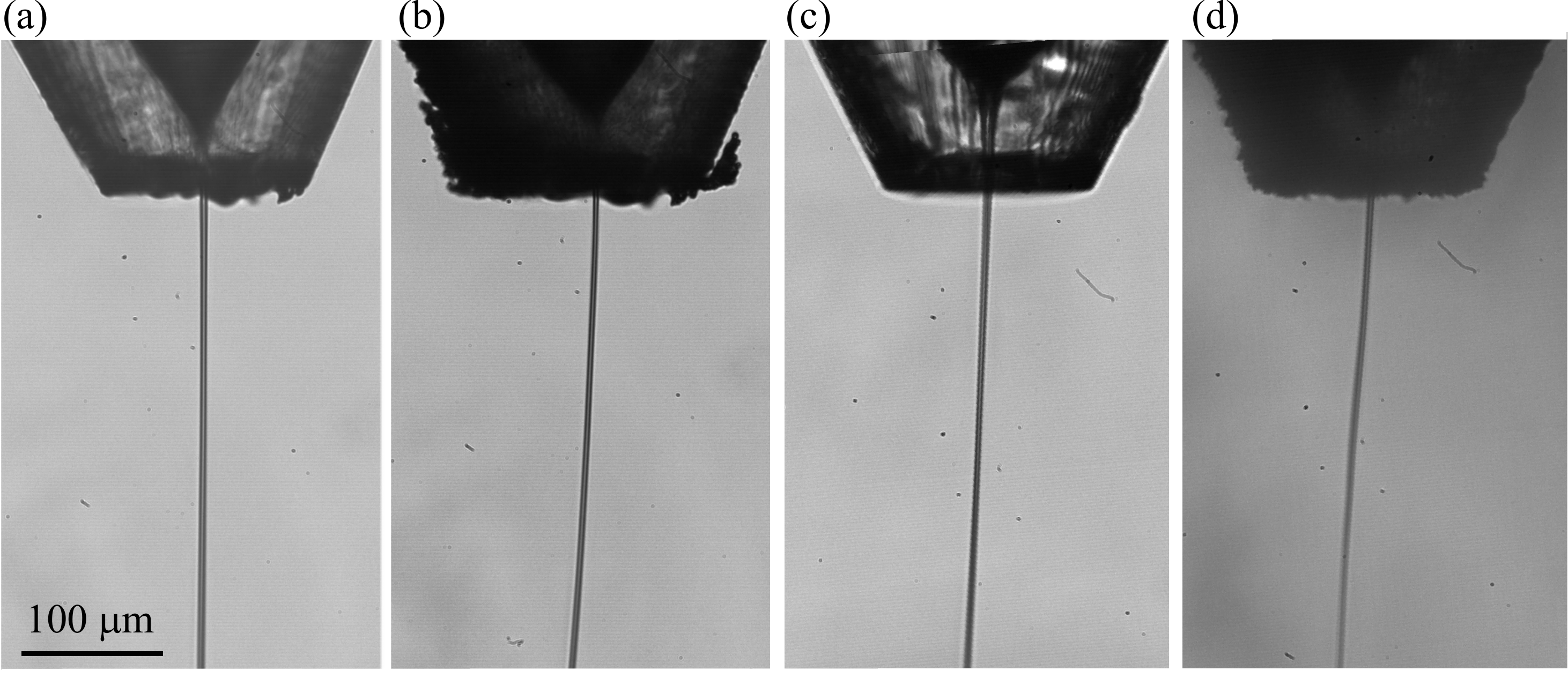}}
\end{center}
\caption{Images of the emitted jet surrounded by a viscoelastic shell of Buffer 1+PEO100K 0.4\% (a),  Buffer 2+PEO100K 0.18\% jet (b), Buffer 3+PEO100K 0.045\% jet (c), and Buffer 4+PEO100K 0.03\% (d). In all cases, the inner liquid was the corresponding buffer fluid. The experiments were conducted for $m'_g=81$ mg/min, $Q_i=5$ $\mu$l/min, and $Q_o=15$ $\mu$l/min.}
\label{image}
\end{figure*}

Hereafter, the PEO100K concentration is the optimum for the corresponding liquid. Figure \ref{image} shows a typical image obtained in the experiments. Interestingly, we did not observe the BOAS instability in any of our experiments, owing to increased shell viscosity and decreased polymer concentration. This may constitute a significant advantage in SFX experiments.

\begin{figure*}[!]
\begin{center}
\resizebox{1.5\columnwidth}{!}{\includegraphics{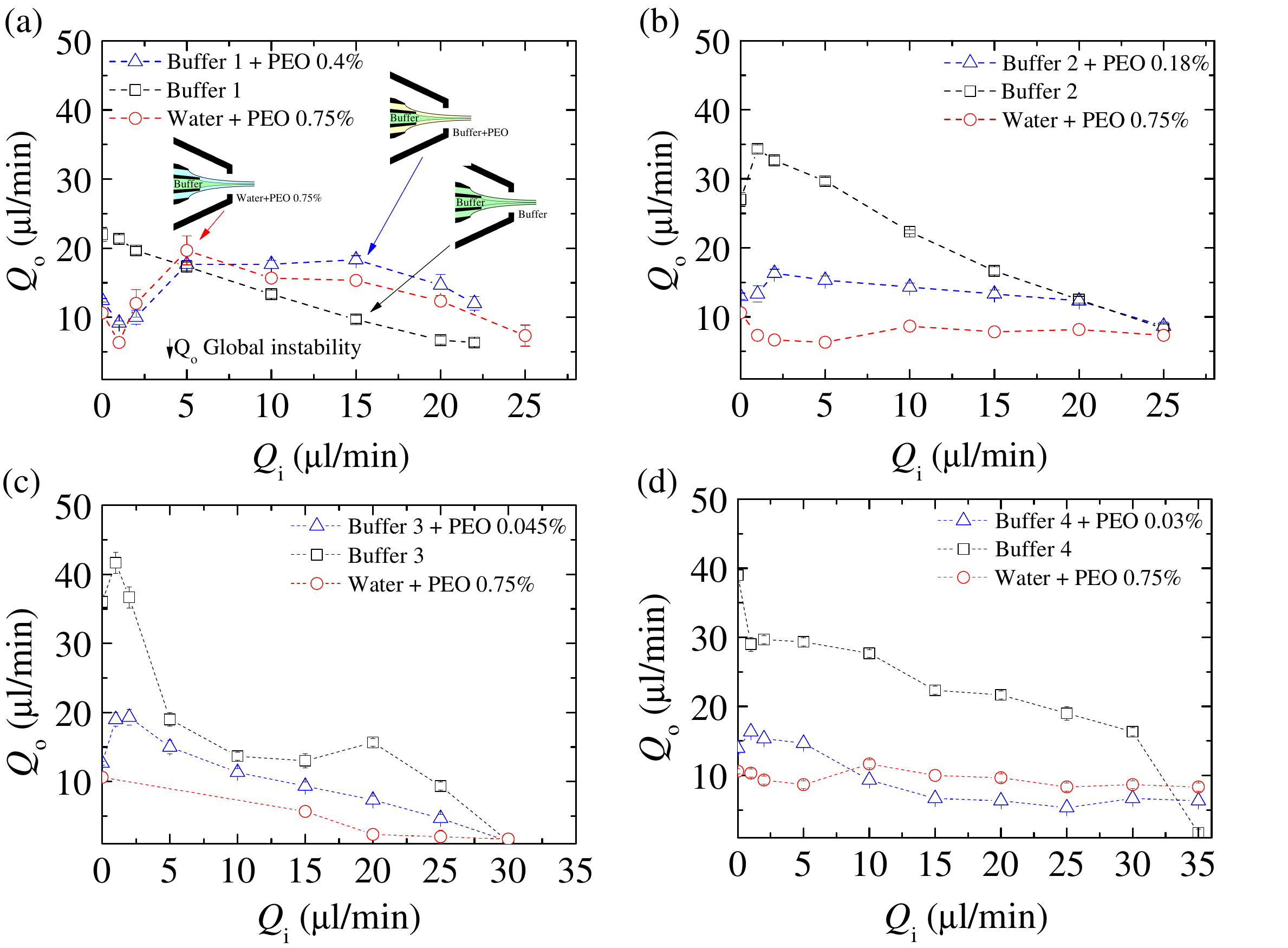}}
\end{center}
\caption{Outer flow rate $Q_o$ as a function of the inner flow rate $Q_i$ at the stability limit for the buffer jet surrounded by (black squares) a shell of the same liquid, (blue triangles) a shell of the same liquid with PEO100K, and (red circles) a shell of an aqueous solution of PEO100K. We do not consider whether there is jet whipping downstream. The experiments were conducted for $m'_g=81$ mg/min.}
\label{limits}
\end{figure*}

We now quantify flow stabilization and its effect on jet diameter and length. We first determined the stability limit for each fluid configuration. In each experimental run, we fixed the inner liquid flow rate $Q_i$ and progressively reduced the outer flow rate $Q_o$ until the stable ejection ceased. We considered a stable ejection if the jet was emitted uninterruptedly for more than 40 s. Figure \ref{limits} shows the outer flow rate $Q_o$ below which the jet emission becomes unstable for a given inner flow rate $Q_i$. Adding PEO100K to Buffer 1 decreases the minimum outer flow rate for the smaller values of $Q_i$ (Fig.\ \ref{limits}a). However, the opposite occurs for larger values of $Q_i$ because the viscoelastic shell triggers a bending (whipping) instability in the tapering meniscus, preventing stable jetting. As \citet{AFMG12} show, this instability exhibits a complex dependence of the parameters governing flow focusing, even in the Newtonian case.

Figures \ref{limits}b and \ref{limits}c show the strong stabilizing effect of the viscoelastic shell at low $Q_i$. In this case, adding a tiny amount of PEO100K can reduce the outer liquid flow rate by a factor of $2-3$. As mentioned above, this effect is explained by the coil-stretch transition \citep{D74} undergone by the polymers in the tapering meniscus. Despite the short polymer relaxation time, the high strain and shear rates generated by the transonic gas stream stretch the polymers under the selected parameters. The elastic stress stabilizes the flow, thereby reducing the minimum flow rate required to achieve the steady jetting regime. For large values of $Q_i$, the ratio $Q_o/Q_i$ is not sufficiently large to produce a significant stabilizing effect. An aqueous viscoelastic shell provides stronger stabilization because of its higher polymer concentration. 

We measured the jet length $L_j$ and diameter $d_j$ at the stability limit. We defined the jet length $L_j$ as the distance from the ejector where the capillary varicose instability arose. We measured the jet diameter $d_j$ at that distance from the ejector. We considered the jet to suffer from whipping (bending) instability if the amplitude of its lateral oscillations exceeded the threshold $0.1\, d_j$. 

\begin{figure*}[!]
\begin{center}
\resizebox{1.5\columnwidth}{!}{\includegraphics{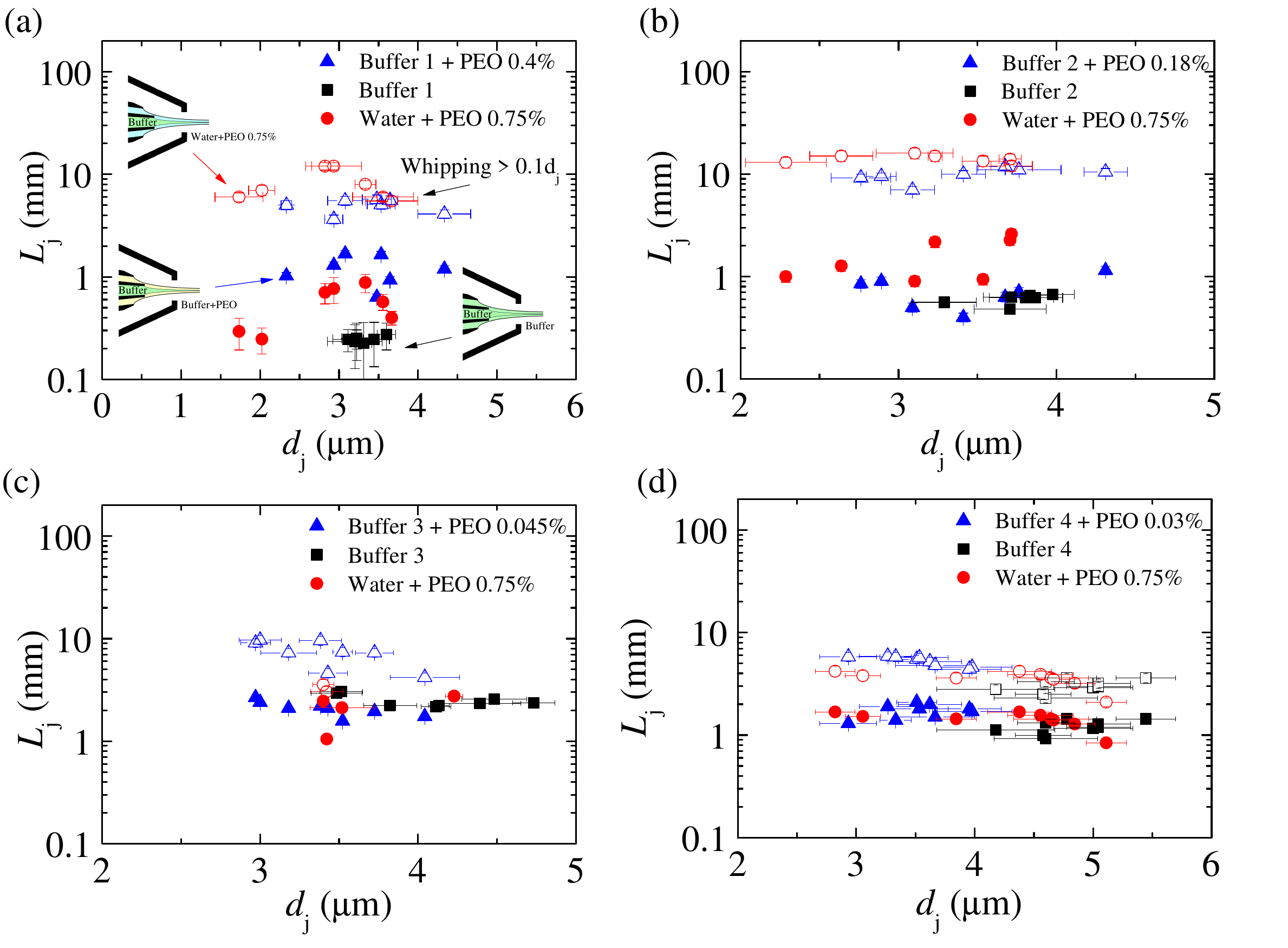}}
\end{center}
\caption{Jet length $L_j$ versus jet diameter $d_j$ at the stability limit for the buffer jet surrounded by (black squares) a shell of the same liquid, (blue triangles) a shell of the same liquid with PEO100K, and (red circles) a shell of an aqueous solution of PEO100K. The open symbols correspond to results without jet whipping (whipping was not observed with a shell of the same liquid). We conducted the experiments at $m'_g=81$ mg/min.}
\label{lengths}
\end{figure*}

\begin{figure*}[!]
\begin{center}
\resizebox{0.7\columnwidth}{!}{\includegraphics{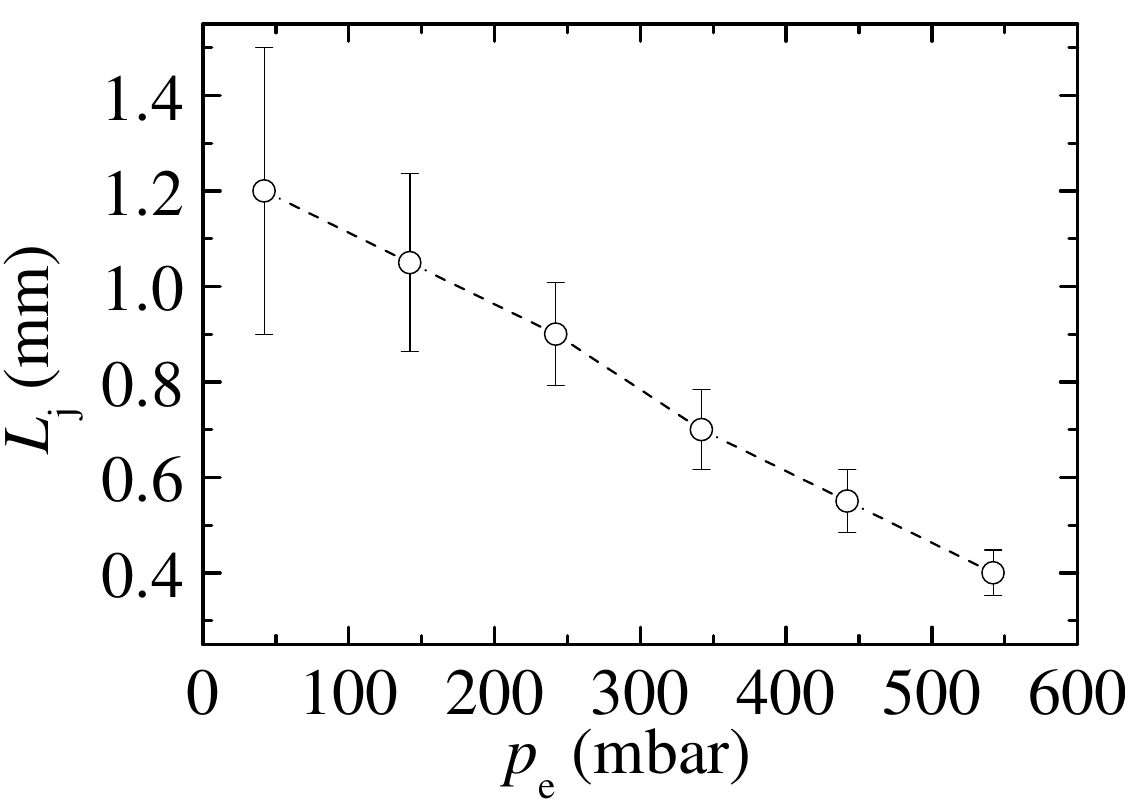}}
\end{center}
\caption{Jet length $L_j$ of the Buffer 1 jet surrounded by a shell of the same liquid with PEO100K without whipping versus exit pressure $p_e$ for $Q_i=15$ $\mu$l/min and $Q_o=18.3$ $\mu$l/min. We conducted the experiments at $m'_g=81$ mg/min.}
\label{whipping}
\end{figure*}

The open squares in Fig.\ \ref{lengths} show the results without considering jet whipping (bending), while the solid squares indicate the length at which whipping occurs. Airstream density strongly enhances this instability. Figure \ref{whipping} shows the length of the Buffer 1 jet without whipping for different exit pressures while keeping the pressure drop $p_0-p_e$ constant. The length without whipping increases as $p_e$ decreases, showing the destabilizing effect of the air density. This implies that whipping may not be a significant limiting factor in SFX, where the jet is discharged in a vacuum.  

The results in Fig.\ \ref{lengths}, without considering jet whipping, show a noticeable positive effect of the thin viscoelastic shell at the limiting flow rates shown in Fig.\ \ref{limits}. Adding PEO100K to the outer liquid produced much thinner and longer jets, even though the polymer's molecular weight and concentration were very small. This effect cannot be attributed to the small increase in the solution viscosity. For instance, in Buffer 1, $\eta_p/\eta=0.22$ while the breakup length increases by two orders of magnitude. Buffer 2 produces micrometer-sized jets more than 1 cm long with just 1800 ppm of PEO100K. The minimum jet diameter decreases from around 5 $\mu$m to 3 $\mu$m. The results in Fig.\ \ref{lengths} show the strong stabilizing effect of the viscoelastic shell even for the Buffer 3 and Buffer 4 jets, despite the tiny PEO100K concentrations. 

\section{Conclusions}

Our experiments have shown the strong stabilizing effect of the viscoelastic shell on flow-focused viscous jets. The results suggest stabilization occurs when $\eta_p$ exceeds a threshold, provided $\lambda$ exceeds the value required for the polymer-coil stretch transition. In this case, the liquid acceleration builds up sufficiently large extensional stresses. Both polymer viscosity and extensional relaxation time increase with shell viscosity, which explains why lower PEO100K concentrations are required to stabilize the viscous jets. Moderately to highly viscous jets are stabilized at very low PEO100K concentrations, at which BOAS instability was not observed. 

Our analysis is limited to a family of solvents (aqueous PEG solutions at different concentrations) and PEO100K. We need to extend the experimental campaign to elucidate several aspects of the problem, including how nozzle geometry and polymer molecular weight affect the optimal polymer concentration for viscous liquids.

\vspace{1cm}

\section*{Acknowledgements}

Support from the Spanish Ministry of Science and Innovation (MCIN) through grant PID2025-171659NB-I00/ AEI// FEDER, UE is gratefully acknowledged. This work was co-funded (85\%) by the European Union through the European Regional Development Fund (FEDER) and the Regional Government of Extremadura. Managing Authority: Ministry of Finance (Grant GR24077). DF-M acknowledges grant PREP2022-000205 funded by MICIU/AEI /10.13039/501100011033 and ESF+. We used generative artificial intelligence tools to revise the text.


\begin{thebibliography}{26}%
\makeatletter
\providecommand \@ifxundefined [1]{%
 \@ifx{#1\undefined}
}%
\providecommand \@ifnum [1]{%
 \ifnum #1\expandafter \@firstoftwo
 \else \expandafter \@secondoftwo
 \fi
}%
\providecommand \@ifx [1]{%
 \ifx #1\expandafter \@firstoftwo
 \else \expandafter \@secondoftwo
 \fi
}%
\providecommand \natexlab [1]{#1}%
\providecommand \enquote  [1]{``#1''}%
\providecommand \bibnamefont  [1]{#1}%
\providecommand \bibfnamefont [1]{#1}%
\providecommand \citenamefont [1]{#1}%
\providecommand \href@noop [0]{\@secondoftwo}%
\providecommand \href [0]{\begingroup \@sanitize@url \@href}%
\providecommand \@href[1]{\@@startlink{#1}\@@href}%
\providecommand \@@href[1]{\endgroup#1\@@endlink}%
\providecommand \@sanitize@url [0]{\catcode `\\12\catcode `\$12\catcode `\&12\catcode `\#12\catcode `\^12\catcode `\_12\catcode `\%12\relax}%
\providecommand \@@startlink[1]{}%
\providecommand \@@endlink[0]{}%
\providecommand \url  [0]{\begingroup\@sanitize@url \@url }%
\providecommand \@url [1]{\endgroup\@href {#1}{\urlprefix }}%
\providecommand \urlprefix  [0]{URL }%
\providecommand \Eprint [0]{\href }%
\providecommand \doibase [0]{http://dx.doi.org/}%
\providecommand \selectlanguage [0]{\@gobble}%
\providecommand \bibinfo  [0]{\@secondoftwo}%
\providecommand \bibfield  [0]{\@secondoftwo}%
\providecommand \translation [1]{[#1]}%
\providecommand \BibitemOpen [0]{}%
\providecommand \bibitemStop [0]{}%
\providecommand \bibitemNoStop [0]{.\EOS\space}%
\providecommand \EOS [0]{\spacefactor3000\relax}%
\providecommand \BibitemShut  [1]{\csname bibitem#1\endcsname}%
\let\auto@bib@innerbib\@empty
\bibitem [{\citenamefont {Ga{\~n}{\'a}n-Calvo}(1998)}]{G98a}%
  \BibitemOpen
  \bibfield  {author} {\bibinfo {author} {\bibfnamefont {A.~M.}\ \bibnamefont {Ga{\~n}{\'a}n-Calvo}},\ }\bibfield  {title} {\enquote {\bibinfo {title} {Generation of steady liquid microthreads and micron-sized monodisperse sprays in gas streams},}\ }\href@noop {} {\bibfield  {journal} {\bibinfo  {journal} {Phys. Rev. Lett.}\ }\textbf {\bibinfo {volume} {80}},\ \bibinfo {pages} {285--288} (\bibinfo {year} {1998})}\BibitemShut {NoStop}%
\bibitem [{\citenamefont {DePonte}\ \emph {et~al.}(2008)\citenamefont {DePonte}, \citenamefont {Weierstall}, \citenamefont {Schmidt}, \citenamefont {Warner}, \citenamefont {Starodub}, \citenamefont {Spence},\ and\ \citenamefont {Doak}}]{DWSWSSD08}%
  \BibitemOpen
  \bibfield  {author} {\bibinfo {author} {\bibfnamefont {D.~P.}\ \bibnamefont {DePonte}}, \bibinfo {author} {\bibfnamefont {U.}~\bibnamefont {Weierstall}}, \bibinfo {author} {\bibfnamefont {K.}~\bibnamefont {Schmidt}}, \bibinfo {author} {\bibfnamefont {J.}~\bibnamefont {Warner}}, \bibinfo {author} {\bibfnamefont {D.}~\bibnamefont {Starodub}}, \bibinfo {author} {\bibfnamefont {J.~C.~H.}\ \bibnamefont {Spence}}, \ and\ \bibinfo {author} {\bibfnamefont {R.~B.}\ \bibnamefont {Doak}},\ }\bibfield  {title} {\enquote {\bibinfo {title} {Gas dynamic virtual nozzle for generation of microscopic droplet streams},}\ }\href@noop {} {\bibfield  {journal} {\bibinfo  {journal} {J. Phys. D: Appl. Phys.}\ }\textbf {\bibinfo {volume} {41}},\ \bibinfo {pages} {195505} (\bibinfo {year} {2008})}\BibitemShut {NoStop}%
\bibitem [{\citenamefont {Montanero}(2024)}]{M24}%
  \BibitemOpen
  \bibfield  {author} {\bibinfo {author} {\bibfnamefont {J.~M.}\ \bibnamefont {Montanero}},\ }\href@noop {} {\emph {\bibinfo {title} {Tip Streaming of Simple and Complex Fluids}}}\ (\bibinfo  {publisher} {Springer Nature},\ \bibinfo {address} {Switzerland},\ \bibinfo {year} {2024})\BibitemShut {NoStop}%
\bibitem [{\citenamefont {{Chapman \em et al.}}(2011)}]{Cetal11}%
  \BibitemOpen
  \bibfield  {author} {\bibinfo {author} {\bibfnamefont {H.~N.}\ \bibnamefont {{Chapman \em et al.}}},\ }\bibfield  {title} {\enquote {\bibinfo {title} {Femtosecond {X}-ray protein nanocrystallography},}\ }\href@noop {} {\bibfield  {journal} {\bibinfo  {journal} {Nature}\ }\textbf {\bibinfo {volume} {470}},\ \bibinfo {pages} {73--79} (\bibinfo {year} {2011})}\BibitemShut {NoStop}%
\bibitem [{\citenamefont {{Wiedorn M. A. \em et al.}}(2018)}]{W18}%
  \BibitemOpen
  \bibfield  {author} {\bibinfo {author} {\bibnamefont {{Wiedorn M. A. \em et al.}}},\ }\bibfield  {title} {\enquote {\bibinfo {title} {Rapid sample delivery for megahertz serial crystallography at {X-ray FELs}},}\ }\href@noop {} {\bibfield  {journal} {\bibinfo  {journal} {IUCrJ}\ }\textbf {\bibinfo {volume} {5}},\ \bibinfo {pages} {574--584} (\bibinfo {year} {2018})}\BibitemShut {NoStop}%
\bibitem [{\citenamefont {Rubio}\ \emph {et~al.}(2022{\natexlab{a}})\citenamefont {Rubio}, \citenamefont {Vega}, \citenamefont {Ga{\~n}\'an-Calvo},\ and\ \citenamefont {Montanero}}]{RVGM22}%
  \BibitemOpen
  \bibfield  {author} {\bibinfo {author} {\bibfnamefont {A.}~\bibnamefont {Rubio}}, \bibinfo {author} {\bibfnamefont {E.~J.}\ \bibnamefont {Vega}}, \bibinfo {author} {\bibfnamefont {A.~M.}\ \bibnamefont {Ga{\~n}\'an-Calvo}}, \ and\ \bibinfo {author} {\bibfnamefont {J.~M.}\ \bibnamefont {Montanero}},\ }\bibfield  {title} {\enquote {\bibinfo {title} {Unexpected stability of micrometer weakly viscoelastic jets},}\ }\href@noop {} {\bibfield  {journal} {\bibinfo  {journal} {Phys. Fluids}\ }\textbf {\bibinfo {volume} {34}},\ \bibinfo {pages} {062014} (\bibinfo {year} {2022}{\natexlab{a}})}\BibitemShut {NoStop}%
\bibitem [{\citenamefont {{M. Vakili \em et al.}}(2026)}]{V26}%
  \BibitemOpen
  \bibfield  {author} {\bibinfo {author} {\bibnamefont {{M. Vakili \em et al.}}},\ }\bibfield  {title} {\enquote {\bibinfo {title} {{PEO-sheathed} liquid jets increase sample delivery stability for serial femtosecond {X-ray} crystallography},}\ }\href@noop {} {\bibfield  {journal} {\bibinfo  {journal} {Sci. Rep.}\ }\textbf {\bibinfo {volume} {16}},\ \bibinfo {pages} {10497} (\bibinfo {year} {2026})}\BibitemShut {NoStop}%
\bibitem [{\citenamefont {Rubio}\ \emph {et~al.}(2022{\natexlab{b}})\citenamefont {Rubio}, \citenamefont {Galindo}, \citenamefont {Vega}, \citenamefont {Montanero},\ and\ \citenamefont {Cabezas}}]{RGVMC22}%
  \BibitemOpen
  \bibfield  {author} {\bibinfo {author} {\bibfnamefont {A.}~\bibnamefont {Rubio}}, \bibinfo {author} {\bibfnamefont {F.}~\bibnamefont {Galindo}}, \bibinfo {author} {\bibfnamefont {E.~J.}\ \bibnamefont {Vega}}, \bibinfo {author} {\bibfnamefont {J.~M.}\ \bibnamefont {Montanero}}, \ and\ \bibinfo {author} {\bibfnamefont {M.~G.}\ \bibnamefont {Cabezas}},\ }\bibfield  {title} {\enquote {\bibinfo {title} {Viscoelastic transition in transonic flow focusing},}\ }\href@noop {} {\bibfield  {journal} {\bibinfo  {journal} {Phys. Rev. Fluids}\ }\textbf {\bibinfo {volume} {7}},\ \bibinfo {pages} {074201} (\bibinfo {year} {2022}{\natexlab{b}})}\BibitemShut {NoStop}%
\bibitem [{\citenamefont {Rubio}\ \emph {et~al.}(2025)\citenamefont {Rubio}, \citenamefont {Montanero}, \citenamefont {Vakili}, \citenamefont {Koua}, \citenamefont {Bajt}, \citenamefont {Chapman},\ and\ \citenamefont {Ga{\~n}{\'a}n-Calvo}}]{RMVKBCG25}%
  \BibitemOpen
  \bibfield  {author} {\bibinfo {author} {\bibfnamefont {A.}~\bibnamefont {Rubio}}, \bibinfo {author} {\bibfnamefont {J.M.}\ \bibnamefont {Montanero}}, \bibinfo {author} {\bibfnamefont {M.}~\bibnamefont {Vakili}}, \bibinfo {author} {\bibfnamefont {F.~H.~M.}\ \bibnamefont {Koua}}, \bibinfo {author} {\bibfnamefont {S.}~\bibnamefont {Bajt}}, \bibinfo {author} {\bibfnamefont {H.~N.}\ \bibnamefont {Chapman}}, \ and\ \bibinfo {author} {\bibfnamefont {A.~M.}\ \bibnamefont {Ga{\~n}{\'a}n-Calvo}},\ }\bibfield  {title} {\enquote {\bibinfo {title} {Superstability of micrometre jets surrounded by a polymeric shell},}\ }\href@noop {} {\bibfield  {journal} {\bibinfo  {journal} {J. Appl. Cryst.}\ }\textbf {\bibinfo {volume} {58 (4)}},\ \bibinfo {pages} {1261--1268} (\bibinfo {year} {2025})}\BibitemShut {NoStop}%
\bibitem [{\citenamefont {Eggers}\ and\ \citenamefont {Villermaux}(2008)}]{EV08}%
  \BibitemOpen
  \bibfield  {author} {\bibinfo {author} {\bibfnamefont {J.}~\bibnamefont {Eggers}}\ and\ \bibinfo {author} {\bibfnamefont {E.}~\bibnamefont {Villermaux}},\ }\bibfield  {title} {\enquote {\bibinfo {title} {Physics of liquid jets},}\ }\href@noop {} {\bibfield  {journal} {\bibinfo  {journal} {Rep. Prog. Phys.}\ }\textbf {\bibinfo {volume} {71}},\ \bibinfo {pages} {036601} (\bibinfo {year} {2008})}\BibitemShut {NoStop}%
\bibitem [{\citenamefont {Entov}\ and\ \citenamefont {Yarin}(1984)}]{EY84}%
  \BibitemOpen
  \bibfield  {author} {\bibinfo {author} {\bibfnamefont {V.~M.}\ \bibnamefont {Entov}}\ and\ \bibinfo {author} {\bibfnamefont {A.~L.}\ \bibnamefont {Yarin}},\ }\bibfield  {title} {\enquote {\bibinfo {title} {The dynamics of thin liquid jets in air},}\ }\href@noop {} {\bibfield  {journal} {\bibinfo  {journal} {J . Fluid Mech.}\ }\textbf {\bibinfo {volume} {140}},\ \bibinfo {pages} {91--111} (\bibinfo {year} {1984})}\BibitemShut {NoStop}%
\bibitem [{\citenamefont {Oliveira}\ and\ \citenamefont {McKinley}(2005)}]{OM05}%
  \BibitemOpen
  \bibfield  {author} {\bibinfo {author} {\bibfnamefont {M.~S.~N.}\ \bibnamefont {Oliveira}}\ and\ \bibinfo {author} {\bibfnamefont {G.~H.}\ \bibnamefont {McKinley}},\ }\bibfield  {title} {\enquote {\bibinfo {title} {Iterated stretching and multiple beads-on-a-string phenomena in dilute solutions of highly-extensible flexible polymers},}\ }\href@noop {} {\bibfield  {journal} {\bibinfo  {journal} {Phys. Fluids}\ }\textbf {\bibinfo {volume} {17}},\ \bibinfo {pages} {071704} (\bibinfo {year} {2005})}\BibitemShut {NoStop}%
\bibitem [{\citenamefont {Blanchet}\ \emph {et~al.}(2023)\citenamefont {Blanchet}, \citenamefont {Round}, \citenamefont {Mertens}, \citenamefont {Ayyer}, \citenamefont {Graewert}, \citenamefont {Awel}, \citenamefont {Franke}, \citenamefont {Dörner}, \citenamefont {Bajt}, \citenamefont {Bean}, \citenamefont {Custódio}, \citenamefont {Juncheng}, \citenamefont {Henkel}, \citenamefont {Gruzinov}, \citenamefont {Jeffries}, \citenamefont {Kim}, \citenamefont {Kirkwood}, \citenamefont {Kloos}, \citenamefont {Knoška}, \citenamefont {Koliyadu}, \citenamefont {Letrun}, \citenamefont {Löw}, \citenamefont {Makroczyova}, \citenamefont {Mall}, \citenamefont {Meijers}, \citenamefont {Murillo}, \citenamefont {Oberthür}, \citenamefont {Round}, \citenamefont {Seuring}, \citenamefont {Sikorski}, \citenamefont {Vagovic}, \citenamefont {Valerio}, \citenamefont {Wollweber}, \citenamefont {Zhuang}, \citenamefont {Schulz}, \citenamefont {Haas}, \citenamefont {Chapman}, \citenamefont {Mancuso},\ and\ \citenamefont
  {Svergun}}]{BRM23}%
  \BibitemOpen
  \bibfield  {author} {\bibinfo {author} {\bibfnamefont {C.~E.}\ \bibnamefont {Blanchet}}, \bibinfo {author} {\bibfnamefont {A.}~\bibnamefont {Round}}, \bibinfo {author} {\bibfnamefont {H.~D.~T.}\ \bibnamefont {Mertens}}, \bibinfo {author} {\bibfnamefont {K.}~\bibnamefont {Ayyer}}, \bibinfo {author} {\bibfnamefont {M.}~\bibnamefont {Graewert}}, \bibinfo {author} {\bibfnamefont {S.}~\bibnamefont {Awel}}, \bibinfo {author} {\bibfnamefont {D.}~\bibnamefont {Franke}}, \bibinfo {author} {\bibfnamefont {K.}~\bibnamefont {Dörner}}, \bibinfo {author} {\bibfnamefont {S.}~\bibnamefont {Bajt}}, \bibinfo {author} {\bibfnamefont {R.}~\bibnamefont {Bean}}, \bibinfo {author} {\bibfnamefont {T.~F.}\ \bibnamefont {Custódio}}, \bibinfo {author} {\bibfnamefont {E.}~\bibnamefont {Juncheng}}, \bibinfo {author} {\bibfnamefont {A.}~\bibnamefont {Henkel}}, \bibinfo {author} {\bibfnamefont {A.}~\bibnamefont {Gruzinov}}, \bibinfo {author} {\bibfnamefont {C.~M.}\ \bibnamefont {Jeffries}}, \bibinfo {author} {\bibfnamefont
  {Y.}~\bibnamefont {Kim}}, \bibinfo {author} {\bibfnamefont {H.}~\bibnamefont {Kirkwood}}, \bibinfo {author} {\bibfnamefont {M.}~\bibnamefont {Kloos}}, \bibinfo {author} {\bibfnamefont {J.}~\bibnamefont {Knoška}}, \bibinfo {author} {\bibfnamefont {J.}~\bibnamefont {Koliyadu}}, \bibinfo {author} {\bibfnamefont {R.}~\bibnamefont {Letrun}}, \bibinfo {author} {\bibfnamefont {C.}~\bibnamefont {Löw}}, \bibinfo {author} {\bibfnamefont {J.}~\bibnamefont {Makroczyova}}, \bibinfo {author} {\bibfnamefont {A.}~\bibnamefont {Mall}}, \bibinfo {author} {\bibfnamefont {R.}~\bibnamefont {Meijers}}, \bibinfo {author} {\bibfnamefont {G.~E.~Pena}\ \bibnamefont {Murillo}}, \bibinfo {author} {\bibfnamefont {D.}~\bibnamefont {Oberthür}}, \bibinfo {author} {\bibfnamefont {E.}~\bibnamefont {Round}}, \bibinfo {author} {\bibfnamefont {C.}~\bibnamefont {Seuring}}, \bibinfo {author} {\bibfnamefont {M.}~\bibnamefont {Sikorski}}, \bibinfo {author} {\bibfnamefont {P.}~\bibnamefont {Vagovic}}, \bibinfo {author} {\bibfnamefont
  {J.}~\bibnamefont {Valerio}}, \bibinfo {author} {\bibfnamefont {T.}~\bibnamefont {Wollweber}}, \bibinfo {author} {\bibfnamefont {Y.}~\bibnamefont {Zhuang}}, \bibinfo {author} {\bibfnamefont {J.}~\bibnamefont {Schulz}}, \bibinfo {author} {\bibfnamefont {H.}~\bibnamefont {Haas}}, \bibinfo {author} {\bibfnamefont {H.~N.}\ \bibnamefont {Chapman}}, \bibinfo {author} {\bibfnamefont {A.~P.}\ \bibnamefont {Mancuso}}, \ and\ \bibinfo {author} {\bibfnamefont {D.}~\bibnamefont {Svergun}},\ }\bibfield  {title} {\enquote {\bibinfo {title} {Form factor determination of biological molecules with x-ray free electron laser small-angle scattering (xfel-sas).}}\ }\href@noop {} {\bibfield  {journal} {\bibinfo  {journal} {Commun. Biol.}\ }\textbf {\bibinfo {volume} {6}},\ \bibinfo {pages} {1057} (\bibinfo {year} {2023})}\BibitemShut {NoStop}%
\bibitem [{\citenamefont {Wollter}\ and\ \citenamefont {Ekeberg}(2024)}]{WE24}%
  \BibitemOpen
  \bibfield  {author} {\bibinfo {author} {\bibfnamefont {A.}~\bibnamefont {Wollter}}\ and\ \bibinfo {author} {\bibfnamefont {T.}~\bibnamefont {Ekeberg}},\ }\bibfield  {title} {\enquote {\bibinfo {title} {Coherent x-ray diffraction imaging of single particles: background impact on 3d reconstruction.}}\ }\href@noop {} {\bibfield  {journal} {\bibinfo  {journal} {J. Appl. Crystallogr.}\ }\textbf {\bibinfo {volume} {57}},\ \bibinfo {pages} {1384--1391} (\bibinfo {year} {2024})}\BibitemShut {NoStop}%
\bibitem [{\citenamefont {Vakili}\ \emph {et~al.}(2026)\citenamefont {Vakili}, \citenamefont {{Rubio Gonz{\'{a}}lez}}, \citenamefont {Sato}, \citenamefont {Kim}, \citenamefont {Wrona}, \citenamefont {E}, \citenamefont {Herrada}, \citenamefont {Bean}, \citenamefont {Montanero},\ and\ \citenamefont {Ga{\~{n}}{\'{a}}n-Calvo}}]{Vakili2026_2}%
  \BibitemOpen
  \bibfield  {author} {\bibinfo {author} {\bibfnamefont {Mohammad}\ \bibnamefont {Vakili}}, \bibinfo {author} {\bibfnamefont {Alejandro}\ \bibnamefont {{Rubio Gonz{\'{a}}lez}}}, \bibinfo {author} {\bibfnamefont {Tokushi}\ \bibnamefont {Sato}}, \bibinfo {author} {\bibfnamefont {Chan}\ \bibnamefont {Kim}}, \bibinfo {author} {\bibfnamefont {Agnieszka}\ \bibnamefont {Wrona}}, \bibinfo {author} {\bibfnamefont {Juncheng}\ \bibnamefont {E}}, \bibinfo {author} {\bibfnamefont {Miguel~A.}\ \bibnamefont {Herrada}}, \bibinfo {author} {\bibfnamefont {Richard}\ \bibnamefont {Bean}}, \bibinfo {author} {\bibfnamefont {Jos{\'{e}}~M.}\ \bibnamefont {Montanero}}, \ and\ \bibinfo {author} {\bibfnamefont {Alfonso~M.}\ \bibnamefont {Ga{\~{n}}{\'{a}}n-Calvo}},\ }\bibfield  {title} {\enquote {\bibinfo {title} {{An XFEL-small-angle scattering study of PEO in solution}},}\ }\href@noop {} {\bibfield  {journal} {\bibinfo  {journal} {in review}\ } (\bibinfo {year} {2026})}\BibitemShut {NoStop}%
\bibitem [{\citenamefont {Rubio}\ \emph {et~al.}(2021)\citenamefont {Rubio}, \citenamefont {Rubio}, \citenamefont {Cabezas}, \citenamefont {Herrada}, \citenamefont {Ga{\~n}{\'a}n-Calvo},\ and\ \citenamefont {Montanero}}]{RRCHGM21}%
  \BibitemOpen
  \bibfield  {author} {\bibinfo {author} {\bibfnamefont {M.}~\bibnamefont {Rubio}}, \bibinfo {author} {\bibfnamefont {A.}~\bibnamefont {Rubio}}, \bibinfo {author} {\bibfnamefont {M.~G.}\ \bibnamefont {Cabezas}}, \bibinfo {author} {\bibfnamefont {M.~A.}\ \bibnamefont {Herrada}}, \bibinfo {author} {\bibfnamefont {A.~M.}\ \bibnamefont {Ga{\~n}{\'a}n-Calvo}}, \ and\ \bibinfo {author} {\bibfnamefont {J.~M.}\ \bibnamefont {Montanero}},\ }\bibfield  {title} {\enquote {\bibinfo {title} {Transonic flow focusing: stability analysis and jet diameter},}\ }\href@noop {} {\bibfield  {journal} {\bibinfo  {journal} {Int. J. Multiphase Flow}\ }\textbf {\bibinfo {volume} {142}},\ \bibinfo {pages} {103720} (\bibinfo {year} {2021})}\BibitemShut {NoStop}%
\bibitem [{\citenamefont {James}(2009)}]{J09}%
  \BibitemOpen
  \bibfield  {author} {\bibinfo {author} {\bibfnamefont {D.~F.}\ \bibnamefont {James}},\ }\bibfield  {title} {\enquote {\bibinfo {title} {Boger fluids},}\ }\href@noop {} {\bibfield  {journal} {\bibinfo  {journal} {Annu. Rev. Fluid Mech.}\ }\textbf {\bibinfo {volume} {41}},\ \bibinfo {pages} {129--142} (\bibinfo {year} {2009})}\BibitemShut {NoStop}%
\bibitem [{\citenamefont {Sousa}\ \emph {et~al.}(2017)\citenamefont {Sousa}, \citenamefont {Vega}, \citenamefont {Sousa}, \citenamefont {Montanero},\ and\ \citenamefont {Alves}}]{SVSMA17}%
  \BibitemOpen
  \bibfield  {author} {\bibinfo {author} {\bibfnamefont {P.~C.}\ \bibnamefont {Sousa}}, \bibinfo {author} {\bibfnamefont {E.~J.}\ \bibnamefont {Vega}}, \bibinfo {author} {\bibfnamefont {R.~G.}\ \bibnamefont {Sousa}}, \bibinfo {author} {\bibfnamefont {J.~M.}\ \bibnamefont {Montanero}}, \ and\ \bibinfo {author} {\bibfnamefont {M.~A.}\ \bibnamefont {Alves}},\ }\bibfield  {title} {\enquote {\bibinfo {title} {Measurement of relaxation times in extensional flow of weakly viscoelastic polymer solutions},}\ }\href@noop {} {\bibfield  {journal} {\bibinfo  {journal} {Rheol. Acta}\ }\textbf {\bibinfo {volume} {56}},\ \bibinfo {pages} {11--20} (\bibinfo {year} {2017})}\BibitemShut {NoStop}%
\bibitem [{\citenamefont {Ferrera}\ \emph {et~al.}(2007)\citenamefont {Ferrera}, \citenamefont {Montanero},\ and\ \citenamefont {Cabezas}}]{FMC07}%
  \BibitemOpen
  \bibfield  {author} {\bibinfo {author} {\bibfnamefont {C.}~\bibnamefont {Ferrera}}, \bibinfo {author} {\bibfnamefont {J.~M.}\ \bibnamefont {Montanero}}, \ and\ \bibinfo {author} {\bibfnamefont {M.~G.}\ \bibnamefont {Cabezas}},\ }\bibfield  {title} {\enquote {\bibinfo {title} {An analysis of the sensitivity of pendant drops and liquid bridges to measure the interfacial tension},}\ }\href@noop {} {\bibfield  {journal} {\bibinfo  {journal} {Meas. Sci. Technol.}\ }\textbf {\bibinfo {volume} {18}},\ \bibinfo {pages} {3713--3723} (\bibinfo {year} {2007})}\BibitemShut {NoStop}%
\bibitem [{\citenamefont {van~der Walt}\ \emph {et~al.}(2012)\citenamefont {van~der Walt}, \citenamefont {Hulsen}, \citenamefont {Bogaerds}, \citenamefont {Meijer},\ and\ \citenamefont {Bulters}}]{WHBMB12}%
  \BibitemOpen
  \bibfield  {author} {\bibinfo {author} {\bibfnamefont {C.}~\bibnamefont {van~der Walt}}, \bibinfo {author} {\bibfnamefont {M.~A.}\ \bibnamefont {Hulsen}}, \bibinfo {author} {\bibfnamefont {A.~C.~B.}\ \bibnamefont {Bogaerds}}, \bibinfo {author} {\bibfnamefont {H.~E.~H.}\ \bibnamefont {Meijer}}, \ and\ \bibinfo {author} {\bibfnamefont {M.~J.~H.}\ \bibnamefont {Bulters}},\ }\bibfield  {title} {\enquote {\bibinfo {title} {Stability of fiber spinning under filament pull-out conditions},}\ }\href@noop {} {\bibfield  {journal} {\bibinfo  {journal} {J. Non-Newtonian Fluid Mech.}\ }\textbf {\bibinfo {volume} {175-176}},\ \bibinfo {pages} {25--37} (\bibinfo {year} {2012})}\BibitemShut {NoStop}%
\bibitem [{\citenamefont {Bird}\ \emph {et~al.}(1987)\citenamefont {Bird}, \citenamefont {C.}, \citenamefont {Armstrong},\ and\ \citenamefont {Hassager}}]{BAH87}%
  \BibitemOpen
  \bibfield  {author} {\bibinfo {author} {\bibfnamefont {R.~B.}\ \bibnamefont {Bird}}, \bibinfo {author} {\bibfnamefont {R.}~\bibnamefont {C.}}, \bibinfo {author} {\bibnamefont {Armstrong}}, \ and\ \bibinfo {author} {\bibfnamefont {O.}~\bibnamefont {Hassager}},\ }\href@noop {} {\emph {\bibinfo {title} {{Dynamics of Polymeric Liquids Volume I: Fluid Mechanics; Volume II: Kinetic Theory}}}}\ (\bibinfo  {publisher} {Wiley},\ \bibinfo {address} {New York},\ \bibinfo {year} {1987})\BibitemShut {NoStop}%
\bibitem [{\citenamefont {Eggers}(1997)}]{E97}%
  \BibitemOpen
  \bibfield  {author} {\bibinfo {author} {\bibfnamefont {J.}~\bibnamefont {Eggers}},\ }\bibfield  {title} {\enquote {\bibinfo {title} {Nonlinear dynamics and breakup of free-surface flows},}\ }\href@noop {} {\bibfield  {journal} {\bibinfo  {journal} {Rev. Mod. Phys.}\ }\textbf {\bibinfo {volume} {69}},\ \bibinfo {pages} {865--929} (\bibinfo {year} {1997})}\BibitemShut {NoStop}%
\bibitem [{\citenamefont {Yarin}\ \emph {et~al.}(2014)\citenamefont {Yarin}, \citenamefont {Pourdeyhimi},\ and\ \citenamefont {Ramakrishna}}]{YPR14}%
  \BibitemOpen
  \bibfield  {author} {\bibinfo {author} {\bibfnamefont {A.~L.}\ \bibnamefont {Yarin}}, \bibinfo {author} {\bibfnamefont {B.}~\bibnamefont {Pourdeyhimi}}, \ and\ \bibinfo {author} {\bibfnamefont {S.}~\bibnamefont {Ramakrishna}},\ }\href@noop {} {\emph {\bibinfo {title} {Fundamentals and Applications of Micro- and Nanofibers}}}\ (\bibinfo  {publisher} {Cambridge University Press},\ \bibinfo {address} {Cambridge, Great Britain},\ \bibinfo {year} {2014})\BibitemShut {NoStop}%
\bibitem [{\citenamefont {Han}\ \emph {et~al.}(2008)\citenamefont {Han}, \citenamefont {Yarin},\ and\ \citenamefont {Reneker}}]{HYR08}%
  \BibitemOpen
  \bibfield  {author} {\bibinfo {author} {\bibfnamefont {T.}~\bibnamefont {Han}}, \bibinfo {author} {\bibfnamefont {A.~L.}\ \bibnamefont {Yarin}}, \ and\ \bibinfo {author} {\bibfnamefont {D.~H.}\ \bibnamefont {Reneker}},\ }\bibfield  {title} {\enquote {\bibinfo {title} {Viscoelastic electrospun jets: {Initial} stresses and elongational rheometry},}\ }\href@noop {} {\bibfield  {journal} {\bibinfo  {journal} {Polymer}\ }\textbf {\bibinfo {volume} {49}},\ \bibinfo {pages} {1651--1658} (\bibinfo {year} {2008})}\BibitemShut {NoStop}%
\bibitem [{\citenamefont {Acero}\ \emph {et~al.}(2012)\citenamefont {Acero}, \citenamefont {Ferrera}, \citenamefont {Montanero},\ and\ \citenamefont {Ga$\tilde{\text{n}}$\'an-Calvo}}]{AFMG12}%
  \BibitemOpen
  \bibfield  {author} {\bibinfo {author} {\bibfnamefont {A.~J.}\ \bibnamefont {Acero}}, \bibinfo {author} {\bibfnamefont {C.}~\bibnamefont {Ferrera}}, \bibinfo {author} {\bibfnamefont {J.~M.}\ \bibnamefont {Montanero}}, \ and\ \bibinfo {author} {\bibfnamefont {A.~M.}\ \bibnamefont {Ga$\tilde{\text{n}}$\'an-Calvo}},\ }\bibfield  {title} {\enquote {\bibinfo {title} {Focusing liquid microjets with nozzles},}\ }\href@noop {} {\bibfield  {journal} {\bibinfo  {journal} {J. Micromech. Microeng.}\ }\textbf {\bibinfo {volume} {22}},\ \bibinfo {pages} {065011} (\bibinfo {year} {2012})}\BibitemShut {NoStop}%
\bibitem [{\citenamefont {Gennes}(1974)}]{D74}%
  \BibitemOpen
  \bibfield  {author} {\bibinfo {author} {\bibfnamefont {P.~G.~De}\ \bibnamefont {Gennes}},\ }\bibfield  {title} {\enquote {\bibinfo {title} {Coil-stretch transition of dilute flexible polymers under ultrahigh velocity gradients},}\ }\href@noop {} {\bibfield  {journal} {\bibinfo  {journal} {J. Chem. Phys.}\ }\textbf {\bibinfo {volume} {60}},\ \bibinfo {pages} {5030} (\bibinfo {year} {1974})}\BibitemShut {NoStop}%
\end{thebibliography}
%

\end{document}